\documentclass[a4paper,fleqn,usenatbib,useAMS]{mnras}

\usepackage{mathptmx}

\usepackage[T1]{fontenc}
\usepackage{ae,aecompl}

\usepackage{graphicx}	
\usepackage{amsmath}	
\usepackage{amssymb}	
\usepackage{multicol}        
\usepackage{bm}		
\usepackage{pdflscape}	

\usepackage{epsfig}
\usepackage{threeparttable} 

\usepackage{epstopdf}
\usepackage[usenames, dvipsnames]{color}

\usepackage{booktabs,array}

\usepackage{color}

\newcommand{\fourU}{4U~1608--52}

\def\lesssim{\mathrel{\hbox{\rlap{\hbox{\lower4pt\hbox{$\sim$}}}\hbox{$<$}}}}
\let\la=\lesssim
\def\gtrsim{\mathrel{\hbox{\rlap{\hbox{\lower4pt\hbox{$\sim$}}}\hbox{$>$}}}}
\let\ga=\gtrsim

\def\deg{\hbox{$^\circ$}}

\def\arcsec{\hbox{$^{\prime\prime}$}}

\def\sol{~\mathrm{M}_\odot}
\def\lx{$L_\mathrm{X}$}
\def\ledd{$L_\mathrm{Edd}$}  

\def\Fx{$F_\mathrm{X}$}

\def\Nh{$N_{\rm H}$}

\def\chired{$\chi^{2}_{\nu}$}
\def\chis{$\chi^{2}$}

\newcommand{\Msun}{\mathrm{M}_{\odot}}
\newcommand{\lum}{\mathrm{erg~s}^{-1}}
\newcommand{\flux}{\mathrm{erg~cm}^{-2}~\mathrm{s}^{-1}}

\newcommand{\nh}{\mathrm{cm}^{-2}}

\newcommand{\rxte}{\textit{RXTE}}

\newcommand{\suzaku}{\textit{Suzaku}}
\newcommand{\maxi}{\textit{MAXI}}

\newcommand{\ergs}{erg s$^{-1}$}

\title[Broad band \suzaku\ spectroscopy of \fourU]{\suzaku\ spectroscopy of the neutron star transient \fourU\ during its outburst decay.}

\author[M. Armas Padilla et al.]{
M. Armas Padilla$^{1,2,3}$\thanks{e-mail: m.armaspadilla@iac.es},
Y. Ueda,$^{3}$,
T. Hori$^{3}$,
M. Shidatsu$^{4}$
and T. Mu\~noz-Darias$^{1,2}$
\\
$^{1}$Instituto de Astrof\'isica de Canarias (IAC), V\'ia L\'actea s/n, La Laguna 38205, S/C de Tenerife, Spain\\
$^{2}$Departamento de Astrof\'isica, Universidad de La Laguna, La Laguna, E-38205, S/C de Tenerife, Spain\\
$^{3}$Department of Astronomy, Kyoto University, Kitashirakawa-Oiwake-cho, Sakyo-ku, Kyoto 606-8502, Japan\\
$^{4}$MAXI Team RIKEN, 2-1 Hirosawa, Wako, Saitama 351-0198, Japan
}

\date{Accepted XXX. Received YYY; in original form ZZZ}

\pubyear{2016}

\begin{document}
\label{firstpage}
\pagerange{\pageref{firstpage}--\pageref{lastpage}}
\maketitle

\begin{abstract}
We test the proposed 3-component spectral model for neutron star low mass X-ray binaries using broad-band X-ray data. We have analysed 4 X-ray spectra (0.8--30~keV) obtained with \suzaku\ during the 2010 outburst of \fourU, which have allowed us to perform a comprehensive spectral study covering all the classical spectral states. We use a thermally Comptonized continuum component to account for the hard emission, as well as two thermal components to constrain the accretion disc and neutron star surface contributions. We find that the proposed combination of multicolor disc, single-temperature black body and Comptonization components successfully reproduces the data from soft to hard states. In the soft state,  our study supports the neutron star surface (or boundary layer) as the dominant source for the Comptonization seed photons yielding the observed weak hard emission, while in the hard state both solutions, either the disc or the neutron star surface, are equally favoured. The obtained spectral parameters as well as the spectral/timing correlations are comparable to those observed in accreting black holes, which support the idea that black hole and neutron star low mass X-ray binaries undergo a similar state evolution during their accretion episodes.
\end{abstract}

\begin{keywords}
accretion, accretion discs -- 
stars: individuals (\fourU) -- 
stars: neutron star -- 
X-rays: binaries
\end{keywords}



\section{Introduction}\label{subsec:intr}

Stellar-mass black holes (BHs) and numerous neutron stars (NSs) are revealed when they form part of low mass X-ray binaries (LMXBs), where they are accreting material from a low mass ($<\sol$) companion star. Accretion takes place via an accretion disc \citep{Shakura1973}, whose inner regions are viscously heated up to $\sim 10^{6-7}$ K, making these sources the brightest in the night X-ray sky.  Depending on the accretion rate, LMXBs can be found in two flavours, transient and persistent sources. The former spend most part of their lives in a faint, quiescent state at X-ray luminosities \lx $\sim 10^{30-32}$ \ergs, but show occasional outburst -- that can last several months -- when brightness increases over a million fold. Persistent systems are always active (i.e. in outburst), with \lx\ $> 10^{36}$ \ergs\ (but see \citealt{ArmasPadilla2013b}).   

Given the rather comparable binary system parameters and gravitational wells, it is not always straightforward to distinguish between NS and BH accretors in LMXBs. A definitive proof for a BH is obtained when the compact object mass exceeds 3$\Msun$ \citep[e.g.][]{Casares2014}, the maximum stable mass of a NS in general relativity \citep{Rhoades1974,Kalogera1996}. In absence of dynamical measurements, NSs are confirmed by events associated with the presence of a solid surface and magnetic field (absent in BHs), such as thermonuclear X-ray burst and coherent pulsations, respectively. From the X-ray point of view, BH-LMXBs are simpler objects with two main spectral components: (i) a hard component believed to arise from inverse-Compton processes in an optically thin inner flow (corona) and typically modelled with a power-law, and (ii) a soft, thermal accretion disc component. The evolving balance between them lead to the presence of the so-called \textit{hard} and \textit{soft} accretion states, which alternate following well-defined hysteresis patterns (e.g. \citealt{Miyamoto1992, Miyamoto1995}) in the spectral and variability domains \citep[][for a review]{Homan2001,Done2007, Munoz-Darias2011, Belloni2011}.

In NS-LMXBs, the X-ray emission from the solid surface and/or boundary layer (hereafter, we will refer this as NS surface for simplicity) is expected to contribute to the X-ray spectrum, in addition to the aforementioned hard (comptonizing corona) and soft (accretion disc) spectral components. This lead to a more complex spectral evolution and state phenomenology \citep{Hasinger1989,VanderKlis2006}. Nevertheless, BH and NS-LMXBs are known to have very similar timing properties \citep[e.g.][]{Wijnands1999}. Recently, BH-like  
hysteresis paths \citep{Munoz-Darias2014} have been found to be a common feature also in NS systems accreting at sub-Eddintong rates (a.k.a. atoll sources).

Since the dawn of X-ray astronomy many efforts have been made to model the NS-LMXBs spectra. In the 80s, two models using differing approaches were proposed, and dubbed \textit{Eastern} \citep{Mitsuda1984,Mitsuda1989} and \textit{Western} \citep{White1988} models. Both include two components, but they differ on their physical origin. On one hand, the \textit{Eastern} model attributes the soft emission to the accretion disc and the hard component to Comptonization of seed photons emitted on the NS surface. In the \textit{Western} model, on the other hand, the NS surface and the Comptonized disc are the sources of the soft and hard components, respectively (see \citealt{Barret2001} for a review). Along the years, both approaches have been successfully applied. As an example, \citet{Done2002}, fitted the same Cyg X-2 data with both methods, obtaining statistically indistinguishable results. Besides model degeneracy, it is not either clear whether any the above methods can be consistently (i.e. keeping the same physical meaning for every component) describe the spectral evolution through the different stages (states) of an outburst. Moreover, given the overall similarity between LMXBs with NS and BH accretors, the large database now available for both groups, and the spectral simplicity expected for the latter, a BH oriented approach could potentially lead to more general conclusions.    

\defcitealias{Lin2007}{LRH07} 
\citet[][hereafter \citetalias{Lin2007}]{Lin2007} tested the two above scenarios and end up proposing a \textit{hybrid}, 3-component model. These are, (i) a constrained broken power-law (CBPL), (ii) a multi-color disc model(MCD) and (iii) a single-temperature blackbody (BB), which account for the Comptonization, the accretion disc and NS surface emissions, respectively. This solution is able to model the spectra of two transient NS-LMXBs (see also \citealt{Lin2009} for an extension) at different states and luminosities, fulfilling physical and phenomenological constraints typically observed in BH systems (see Section \ref{subsec:test}). Likewise, by combining this modelling with the evolution of the fast temporal variability, a consistent BH-NS accretion state picture is found even for the brightest NS-LMXBs (a.k.a. Z sources). This includes similar accretion-outflow coupling properties for both families \citep[][for a review]{Munoz-Darias2014, Fender2016}.
 
One of the limitations of the \citetalias{Lin2007} study is the lack of coverage at energies $\lesssim 3$ keV (they used \rxte\ data), which hampered a detailed modelling of the thermal components (disc and NS surface). This is particularly troublesome in the hard state, when these are weaker. As a matter of fact, \citetalias{Lin2007} found that the accretion disc component was not required to fit the hard state observations, a limitation commonly suffered in BH systems when using the same instrumentation \citep[e.g.,][]{Munoz-Darias2013}. However, hard state discs are detected in the same BHs when observing with instruments providing lower energy coverage \citep[e.g.][]{Plant2015, DeMarco2015}.  

In this paper we take the \citetalias{Lin2007} approach as the starting point to analyse 4 broad-band (0.8--30 keV) \suzaku\ observations covering the 2010 outburst decay of \fourU\ from soft to hard states. This source, one of the two used to test the aforementioned spectral model, is a well-known transient NS-LMXB \citep{Grindlay1976,Tananbaum1976}, which recurrently undergoes into outburst \citep[e.g.][]{Negoro2014}. From radius-expansion type I X-ray burst the distance is estimated to be in the range 3.2--4.1~kpc \citep[][]{Nakamura1989,Galloway2008}, and the detection of rapid oscillation in some of these events indicates that the NS spins at 620 Hz \citep{Muno2001}. \fourU\ has been extensively studied, with works published on its X-ray variability properties \citep[e.g.][]{Yoshida1993,Altamirano2008,Barret2013}, burst episodes \citep[e.g.][]{Galloway2008,Poutanen2014}, disc reflection features  \citep[e.g.][]{Degenaar2015} and spectral properties \citep[e.g.][\citetalias{Lin2007}]{Gierlinski2002, Asai2012}. The latest group, however, was mostly carried out with data from instruments with a coverage-lack at low energies, which results in strong limitations when modelling the soft components; a caveat that can be accounted for by using \suzaku\ data. \\

\section{Observations and data reduction}\label{subsec:obs}
\begin{figure}
\begin{center}
\includegraphics[keepaspectratio,width=\columnwidth]{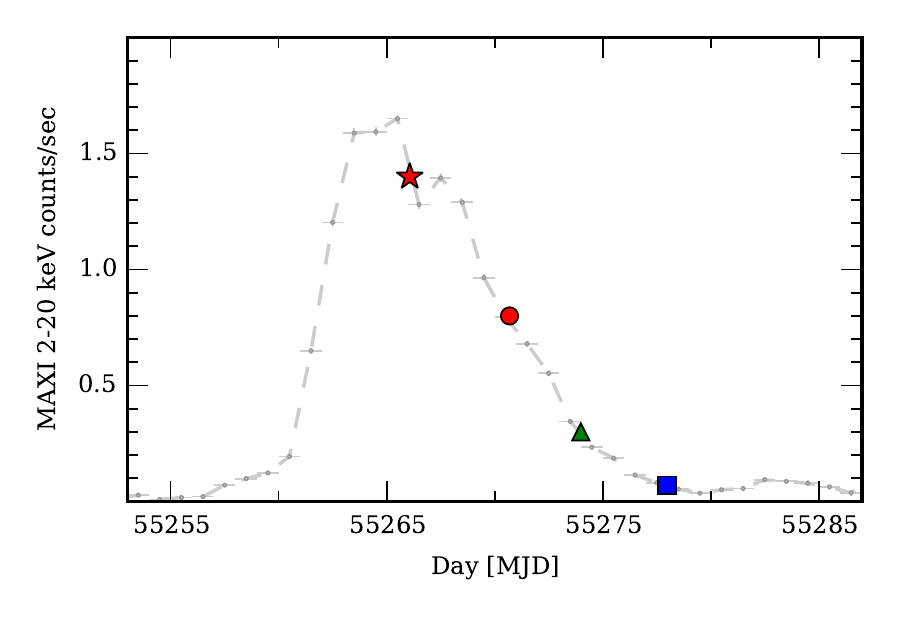}\\
\caption{The 2010 outburst light curve (2-20 keV) of \fourU\ obtained by \maxi . The times of the 4 \suzaku\ observations are indicated by the coloured symbols. Observation 1, 2, 3, and 4 are  indicated by a red star, a red circle  and a blue square, respectively. Same symbol/colour code is used for all the figures.}
\label{fig:LC}
\end{center}
\end{figure}

\fourU\ was observed with the spacial facility \suzaku\ \citep{Mitsuda2007} during its 2010 outbourst. Four observations were performed throughout the outburst decay, on 2010 March 11, 15, 18 and 22, for a total observing time of $\sim$114~ksec (see Figure \ref{fig:LC}).  We used the \textsc{ heasoft} v.6.18 software and \suzaku\ Calibration Database (CALDB) released on 2015 March 12 for the reduction and analysis of our data. After reprocessing the data with the \suzaku\ {\ttfamily FTOOL}  {\ttfamily aepipeline} and removing burst events present in observations 404044020 and 404044030, we followed the \suzaku\ ABC guide\footnote{\url{http://heasarc.gsfc.nasa.gov/docs/suzaku/analysis/abc/}} to obtain the final data products.

\begin{table*}
\centering
\caption{\suzaku\ observations log.}
\begin{threeparttable}
\begin{tabular}{c c c  c c c c c}
\hline
\multicolumn{3}{c}{Observations}  & \multicolumn{2}{c}{Date} & \multicolumn{2}{c}{XIS}& HXD-PIN\\
 (Obs) &  & ID &MJD& yyyy-mm-dd &Burst option  & Net exposure$^{a}$ & Net exposure$^{a}$\\
\hline
1 & \textcolor{red}{$\bigstar$} & 404044010 & 55266.07 &2010-03-11  & 0.13 sec  & 2.2 ksec & 28 ksec \\
2 & \textcolor{red}{$\bullet$} & 404044020 & 55270.69 &2010-03-15 & 0.13 sec  & 2.3 ksec & 25.1 ksec \\
3 & \textcolor{ForestGreen}{$\blacktriangle$} & 404044030 & 55273.99 &2010-03-18  & 0.5 sec  & 7.8 ksec & 14.1 ksec \\
4 & \textcolor{blue}{$\blacksquare$} & 404044040 & 55277.98 &2010-03-22  & 1.0 sec  & 15.9 ksec & 14.5 ksec \\

\hline
\end{tabular}
\begin{tablenotes}
\item[a]{Exposure time per sensor after dead-time and burst events substration.}
\end{tablenotes}
\label{tab:obs}
\end{threeparttable}
\end{table*}

\begin{figure*}
\begin{center}
\includegraphics[keepaspectratio]{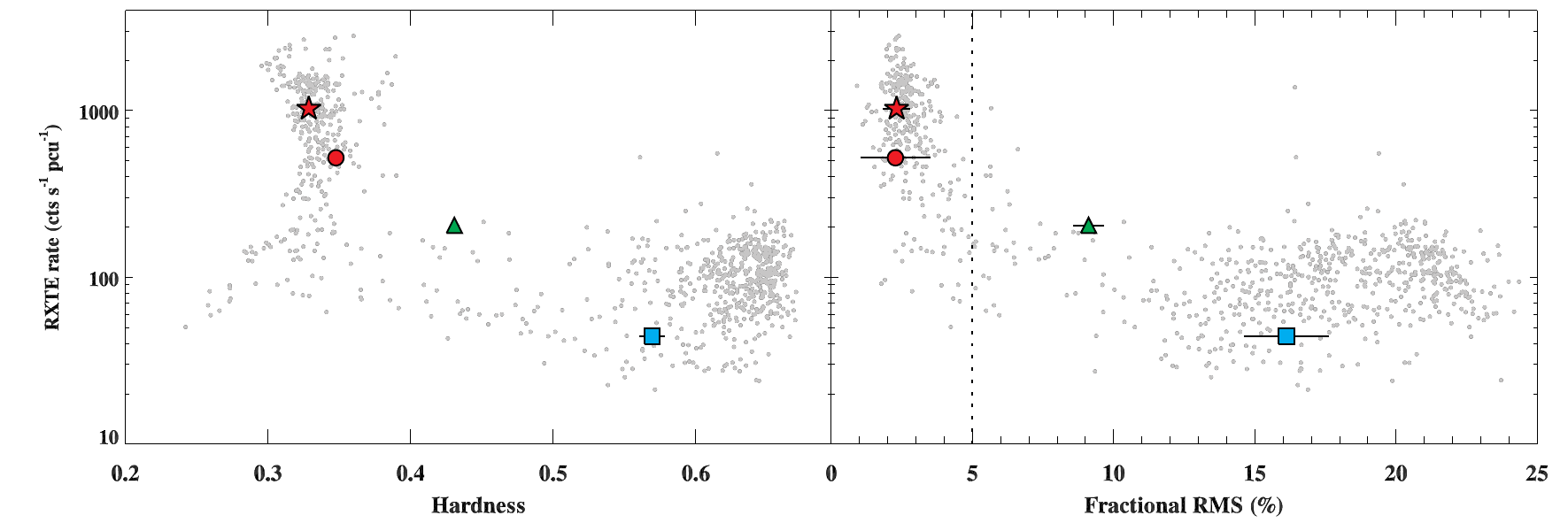}\\
\caption{Hardness intensity diagram (left panel) and rms intensity diagram (right panel) of \fourU\ using RXTE observations (see text). Data simultaneous to our \suzaku\ campaign  are highlighted using the same colour scheme as in Fig. \ref{fig:LC}. The Dashed line represents the 5 \% variability level, which roughly define the soft states in BH and NS LMXBs. Adapted from \citet{Munoz-Darias2014}.}
\label{fig:RXTE}
\end{center}
\end{figure*}


The X-ray Imaging Spectrometer (XIS; \citealt{Koyama2007}) was operated in the 1/4 window mode and using the burst option to avoid pile-up effects (see Table\ref{tab:obs} for the observation log). We used a circular region with a radius of 110\arcsec centred on the system position to extract the source events, and a circular region of the same size placed in a source-free part of the CCDs to extract the background. The XIS response matrix and ancillary response files were created using {\ttfamily xisrmfgen} and {\ttfamily xisarfgen} tasks.
We made use of the {\ttfamily aepileupcheckup.py} script developed by \citet{Yamada2012} to evaluate the pile-up fraction in our observations. It returned an estimation of less than 2\% at the core of the PSFs. Nevertheless, to ensure that our spectral results are not affected by pile-up, we extracted the source spectra using several annular regions excluding photons coming from the most central region. The spectra are consistent with each other confirming that the pile-up effect is negligible. 
We combined spectra and response files of the two frontside-illuminated cameras (FI-XISs; XIS-0 and XIS-3) with the {\ttfamily addascaspec} task to maximize the signal-to-noise ratio. The spectra from the backside-illuminated camera XIS-1 was excluded from our study due to discrepancies (probably by inaccurate cross calibration) with the FI-XISs, which are more sensitive at high energies.

We extracted the spectra from PIN silicon diodes of the Hard X-ray Detector (HXD; \citealt{Takahashi2007}) using the {\ttfamily FTOOL} {\ttfamily hxdpinxbpi} provided by the \suzaku\ team. This script produces the background spectrum extracting the non-X-ray background (NXB) from the tunned background modeled by the HXD team, and adding the contribution of the cosmic X-ray background (CXB) following the \citet{Boldt1987} recipe. Due to the Galactic coordinates of the source (\textit{l}=330.92\degr, \textit{b}=-0.85\degr), we also account for contamination caused by Galactic ridge X-ray emission  following \citet{Revnivtsev2003,Revnivtsev2006}.
In our analysis we used the appropiate response file provided on the \suzaku\ CALDB. The source was not significantly detected with the Gadolinium Silicate (GSO) scintillator.
 
\subsection{Rossi X-ray Timing Explorer}\label{subsec:RXTE} 

The four \suzaku\ observations were (at least) partially covered by simultaneous \textit{Rossi X-ray Timing Explorer} (RXTE) data. In Fig. \ref{fig:RXTE} we present both the hardness-intensity and rms-intensity diagrams (HID and RID, respectively) from \citet{Munoz-Darias2014}. These include every (980) RXTE pointing of \fourU\ during its $\sim 15$ years of operations. Every RXTE observation is represented by a grey dot.  Observations simultaneous with the \suzaku\ data are highlighted using the same colour code as in Fig. \ref{fig:LC}. The RID was performed following \citet{Munoz-Darias2011} by computing fractional root-mean-square (rms) in the 0.1--64 Hz frequency band. In the HID, \textit{Hardness} was defined as the ratio of counts between the 10--16 keV and 6--10 keV bands, respectively. RXTE net count-rate (Y axis in Fig. \ref{fig:RXTE}) is computed in the band 2--15 keV. Hard states are found at rms $\gtrsim 15$ \% and hardness $\gtrsim 0.6$, whilst soft states are characterized by rms $\lesssim 5$ \% and hardness $\lesssim 0.4$ \citep[see][for further details]{Munoz-Darias2014}.
 
\section{Analysis and results}\label{sec:results}

We used \textsc{xspec} (v.12.9, \citealt{Arnaud1996}) to analyse our spectra. 
We added a 1 per cent systematic error to every spectrum to account for calibration uncertainties \citep{Makishima2008}, and excluded both the 1.7-1.9~keV and 2.2-2.4~keV energy ranges to avoid the silicon K-edge and gold M-edge calibration uncertainties, respectively.
In order  to deal with cross-calibration issues we added a constant factor (CONSTANT) to the spectral models with a value fixed to 1 for XIS spectra and 1.16 for HXD spectra \footnote{Suzaku Memo 2008-06 at \url{http://www.astro.isas.jaxa.jp/suzaku/doc/suzakumemo}}. 
We included the photoelectric absorption component (PHABS) to account for the interstellar absorption assuming the cross-sections of \citet{Balucinska-Church1992} and the abundances of \citet{Anders1989}. Finally, the 1-10~keV FI-XISs and 15-30~keV HXD/PIN spectra were simultaneously fitted by using tied spectral parameters. We note that HXD/PIN data above 30~keV were excluded since they do not exceed the background level (at 3$\sigma$).\footnote{The 40-70~keV NXB have a 3$\sigma$ systematic error of 8.4-5.4$\%$ for an exposure of 10-20~ksec, respectively \citep{Fukazawa2009}} \par 
  
In this paper we aim at modelling 4 \suzaku\ spectra using the \citetalias{Lin2007} 3-component model. The broader energy range (particularly in the soft band) of our data allow us to give an step forward with respect to previous works. On one hand, and differently from \citetalias{Lin2007} who used a (constrained) broken power-law to model the hard component, we utilize a more physically meaningful thermally Comptonized continuum component (NTHCOMP in \textsc{xspec}; \citealt{Zdziarski1996}; \citealt{Zycki1999}). On the other hand, we apply the very same 3-component model also to the hard state spectrum (obs. 4), contrastingly to the simpler 2-component approach proposed in the hybrid model. Thus, we use the model DISKBB+BBODYRAD+NTHCOMP in \textsc{xspec} to fit soft (obs. 1 and 2), intermediate (obs. 3) and hard (obs. 4) state data of the same source taken during the same outburst. Results are reported in Table \ref{tab:res} and  Fig. \ref{fig:fit} with uncertainties given at 90 per cent confidence level. 

We have two possible seed-photon solutions for the Comptonized emission. One assumes that the up-scattered photons arise from the disc and the second associates them with the NS surface. Thus, we coupled $KT_{\rm seed}$ to either $KT_{\rm disk}$ or $KT_{\rm bb}$, respectively, and changed the seed photons shape parameter accordingly ($im\_type$ in NTHCOMP; see Table \ref{tab:res}). Both approaches yield statistically indistinguishable results (\chired$<$1.05). In Table \ref{tab:res}, we report only the case in which the seed photons arise from the NS surface (see Section \ref{sec:Dissc} for discussion). On the other hand, given the complexity of the modelling, uncertainties have been calculated in two steps, that is, freezing the thermal components when calculating the error associated with the hard tail (NTHCOMP) and vice versa.

We note that, in contrast to previous studies of the low state spectra where a reflection component was required (e.g, \citealt{Yoshida1993} using \textit{Ginga} data; \citealt{Degenaar2015} using \textit{NuSTAR}), our modelling do not require reflection related features. As a test, we added to our models a Gaussian component in order to search for the presence of a Fe-K line with central energy of 6.4~keV and a width of 0.1~keV, consistently with previous works (e.g. \citetalias{Lin2007}). We always found upper limits as stringent as Fe-K equivalent width $<$~14eV. Therefore, the reflection features from a cold outer disk are either absent or buried by the continuum.

\subsection{Spectral parameters}

Spectral fits yield a hydrogen column density (\Nh) of $\sim 1\times 10^{22}\nh$, which agrees with values previously reported \citep{Penninx1989,Guver2010}. The inferred 0.8--30~keV unabsorbed flux goes from 13.8$\times10^{-9}\flux$ in the soft state, to 0.6 $\times10^{-9}\flux$ in the hard state. Assuming a distance of 3.6~kpc, they translate into a luminosity drop from 21.5 to 0.9 $\times10^{36}\lum$ (0.1 to 0.004~\ledd\ assuming an Eddington luminosity  of \ledd=2$\times10^{38}\lum$). The soft-to-hard transition occurs at $\sim 0.02$~\ledd, which agrees with the 0.01-0.04~\ledd\ range proposed by \citet{Maccarone2003} for this transition in BH and NS LMXBs. 

The disc and NS temperatures behave as expected for the corresponding accretion rates, since they drop from 1 to 0.2~keV and from 1.6 to 0.4~keV, respectively. From the black body normalization we infer an emission radius ($R_{\rm bb}$) in the range of $\sim$1--6~km, which is consistent with previous works \citep[e.g.,][]{Sakurai2014}. These small $R_{\rm bb}$ values can be interpreted as resulting from emission from a boundary layer taking the shape of an equatorial belt (\citetalias{Lin2007} and references therein; see also \citealt{Matsuoka2013}). We note, however, that it is not straightforward to obtain physically meaningful $R_{\rm bb}$ values from this kind of modelling. Also, we did not apply any correction for the photons scattered by the Corona which might increase the obtained values by a factor $<$2 \citep[see ][]{Kubota2004}.  

Finally, the (Comptonization) asymptotic power-law photon index ($\Gamma$) slightly decreases as the flux goes down, from $\sim$2.3 to 2, whereas the electron corona temperature ($KT_{\rm e}$) increases from 3.3 to 18~keV. These values correspond to a decrease on the electron scattering optical depth ($\tau$) from $\sim$ 8 to  $\sim$3 \footnote{Photon index is related to $kT_{e}$ and optical depth ($\tau$) through the relation $\Gamma=\alpha +1=\left[\frac{9}{4}+\frac{1}{(kT_{e}/m_{e}c^{2}) \tau (1+ \tau /3)} \right]^{1/2}-\frac{1}{2}$ \citep{Sunyaev1980,Lightman1987}}. The thermal components contribute by $\sim 77$ per cent to the total luminosity in the soft state ($\sim 54$ per cent from the disc and 23 per cent from the NS surface). On the contrary, the thermal contribution is only $\sim$15 per cent in the hard state .\\

\subsection{Some initial sanity tests}\label{subsec:test}

\citetalias{Lin2007} approached model degeneracy by setting a number of physical evaluation criteria. These are:\\
1. The inner disc radii have values comparable/higher than the size of the NS. \\
2. The thermal components (MCD and BB) roughly evolve as \lx$\propto T^{4}$.\\
3. The disc temperature is lower than the NS surface temperature \citep[$KT_{\rm diskbb} < KT_{\rm bb}$;][]{Mitsuda1989,Popham2001}; and therefore the seed photons temperature should not exceed the NS surface temperature ($KT_{\rm seed} \leq KT_{\rm bb}$). \\
4. The Comptonization fraction is consistent with the power density spectrum, assuming that in NS-LMXBs the X-ray variability changes as a function of spectral hardness similarly to the BHs systems. \\

In order to test the first condition we  converted the normalization of the disc component ($N_{\rm diskbb}$) into inner disk radius ($R_{\rm in}$) following the relation: 

\begin{equation}
R_{\rm in}=\xi \kappa^{2} (\frac{N_{\rm disk}}{cos~i})^{0.5} \frac{D}{10~ \rm kpc} ~\rm [km]
\label{eq:Rin}
\end{equation}

\citep{Kubota1998,Gierlinski2002}. \textit{D} is the source distance, \textit{i} is the disc inclination angle, $\kappa$ is the ratio of color temperature to effective temperature \citep{Shimura1995}, and $\xi$ is the correction factor for the inner torque-free boundary condition \citep{Kubota1998}, although we note that the torque-free boundary condition is most likely not fulfilled in NS-LMXBs. Here we adopt $\kappa$=1.7 ,  $\xi$=0.4 and $i= 70$\deg. The obtained inner disc radii  are in the range  $R_{\rm in}\sim$15--45~km. These values become slightly smaller ($R_{\rm in}\sim$11--30~km) if  $i=40$\deg \citep{Degenaar2015}, but are in any case larger than the expected NS radius \citep[e.g.][]{Lattimer2000,Zdunik2012}. These values are consistent with those obtained by \citet{Degenaar2015}, which use a reflection model to fit a  \fourU\ \textit{NuSTAR} observation at $\sim$0.02 \ledd\ (i.e. similar to those consider here).

Condition 3 is also fulfilled (see Table \ref{tab:res}), while Condition 4 is discussed in Section \ref{sec:Dissc} since it has important implications on the nature of the Comptonized emission. Finally, Condition 2 only makes sense if the inner disc radius is constant with time. We observe than both the black-body and the disc components roughly evolved as \lx$\propto T^{4}$\ (not showed), which is consistent with the rather small variation measured in the inner radius.  Nevertheless, the reader should bear in mind that our disc radius measurements are derived from a Newtonian model (DISKBB) and should be taken with caution when used for more complex calculations.

\begin{table*}
\centering
\caption{Fitting results for the DISKBB+BBODYRAD+NTHCOMP model for the 4 observations assuming that seed-photons arise from the NS surface ($KT_{\rm seed}$ tied to $KT_{\rm bb}$). The fifth column shows the results for obs.4 with the same model but assuming that seed-photons arise from the disc ($KT_{\rm seed}$ tied to $KT_{\rm in}$). Last column shows the results for obs. 4 using the simpler  DISKBB+NTHCOMP model ($KT_{\rm seed}$ tied to $KT_{\rm in}$). Both spectra and models are shown in Fig. \ref{fig:fit}. Uncertainties are expressed at 90 per cent confidence level.
}
\begin{threeparttable}
\begin{tabular}{ l c c c c c c }
\hline
Component  & Obs1 (SS) & Obs2 (SS) &Obs3 (IS) & Obs4 (HS) &\multicolumn{2}{c}{Obs4 (HS) alternative models}\\
 		 & 		 & 		 	&		 		& 				 &	seed photons (\textsc{diskbb})	& (\textsc{diskbb+nthcomp})\\
\hline
\Nh\ ($\times 10^{22}\nh$)	& 	1.07$\pm$0.01 			& 	1.05$\pm$0.01 	&  1.01$\pm$0.02			&	0.99$\pm$0.15  		& 1.1$\pm$0.1 		& 	1 (fix)		\\
$KT_{\rm in}$ (keV)			& 	1.00$\pm$0.006			& 	0.92$\pm$0.006	&  0.621$\pm$0.004 		&   0.23	$\pm$0.02		& 0.15$\pm$0.01  	& 	0.40$\pm$0.02 \\
$N_{\rm diskbb}$				&   448$\pm$10				& 	386$\pm$10 		&  623$\pm$17			& 	4160 $^{+2046}_{-1346}$&53694$^{+70658}_{-25350}$ &140$^{+27}_{-13}$ \\
$KT_{\rm bb}$ (keV)			&   	1.63$\pm$0.004 			& 	1.70$\pm$0.006 	&  1.25$\pm$0.005		&	0.395$\pm$0.001 		& 0.42$\pm$0.01 		&	 --	 				\\
$N_{\rm bb}$					&  40$\pm$1.2				& 	21.2$\pm$0.7		&  14.7$\pm$0.6 			&	277$\pm$	27			& 143$\pm$22 		&	-- 				\\
$\Gamma$/$\tau$ $^{a}$ 		& 	2.26$\pm$0.06/7.69		& 	2.18$\pm$0.1/7.8& 	2.14$\pm$0.03/3.53	&	1.98$\pm$0.01/3.23	& 1.98$\pm$0.01/3.15 &	2.03$\pm$0.02  \\
$KT_{\rm e}$ (keV)			& 	3.31$\pm$0.09			& 	3.2	$\pm$0.1 	& 	14.5$^{+6.5}_{-2.9}$	&	18$^{+14}_{-4}$		& 19$^{+14}_{-5}$ 	&	999$^{+29}_{-5}$	\\
imp$\_$type					& 	0						&   0				& 0						&   0					& 1					& 1			\\
$N_{\rm nthcomp}$ ($\times 10^{-2}$) & 	3.16$\pm$0.06   	& 	1.44	$\pm$0.04	& 1.71$\pm$0.02   		&	5.11	$\pm$0.04		& 12.3$\pm$0.1  		&	9.8$\pm$0.7			\\
\Fx$^{b}$ ($\flux$)			&    14.20$\pm$0.06			&	8.33$\pm$0.05	& 3.09$\pm$0.03			&	0.76$\pm$0.02		& 0.83$\pm$0.04 		&	0.76$\pm$0.02		\\
\lx$^{c}$ ($\lum$)			&    2.20$\pm$0.01			&	1.29	$\pm$0.01	& 0.479$\pm$0.005		&	0.118$\pm$0.003		& 0.128$\pm$0.006 	&	0.118$\pm$0.003		\\
MCD fraction	 				&	54						&	57				&	41					&	7					& 3 					&	7					\\
BB fraction					&	23						&	23				&	12					&	8					& 5 					&	--					\\
NTHCOMP fraction				&	23						&	20				&	47					&	85					& 92					&	93			\\
\chired\ /dof 				& 	0.96/1859 				& 	1.02/1324 		& 1.03/1370 				&	1.03/1400			& 1.03/1400  		&	1.06/1403			\\

\hline
\end{tabular}
\begin{tablenotes}

\item[a]{The electron scattering optical depth ($\tau$) is obtained following the relation $\Gamma=\left[\frac{9}{4}+\frac{1}{(kT_{e}/m_{e}c^{2}) \tau (1+ \tau /3)} \right]^{1/2}-\frac{1}{2}$   }
\item[b]{Unabsorbed 0.8-30~keV flux in units of 10$^{-9}\flux$.}
\item[c]{0.8-30~keV luminosity in units of 10$^{37}\lum$ assuming a distance of D=3.6~kpc.}

\end{tablenotes}
\label{tab:res}
\end{threeparttable}
\end{table*}

\begin{figure*}
\begin{center}
\includegraphics[keepaspectratio,trim=0.5cm 0.5cm 0.5cm 1.0cm]{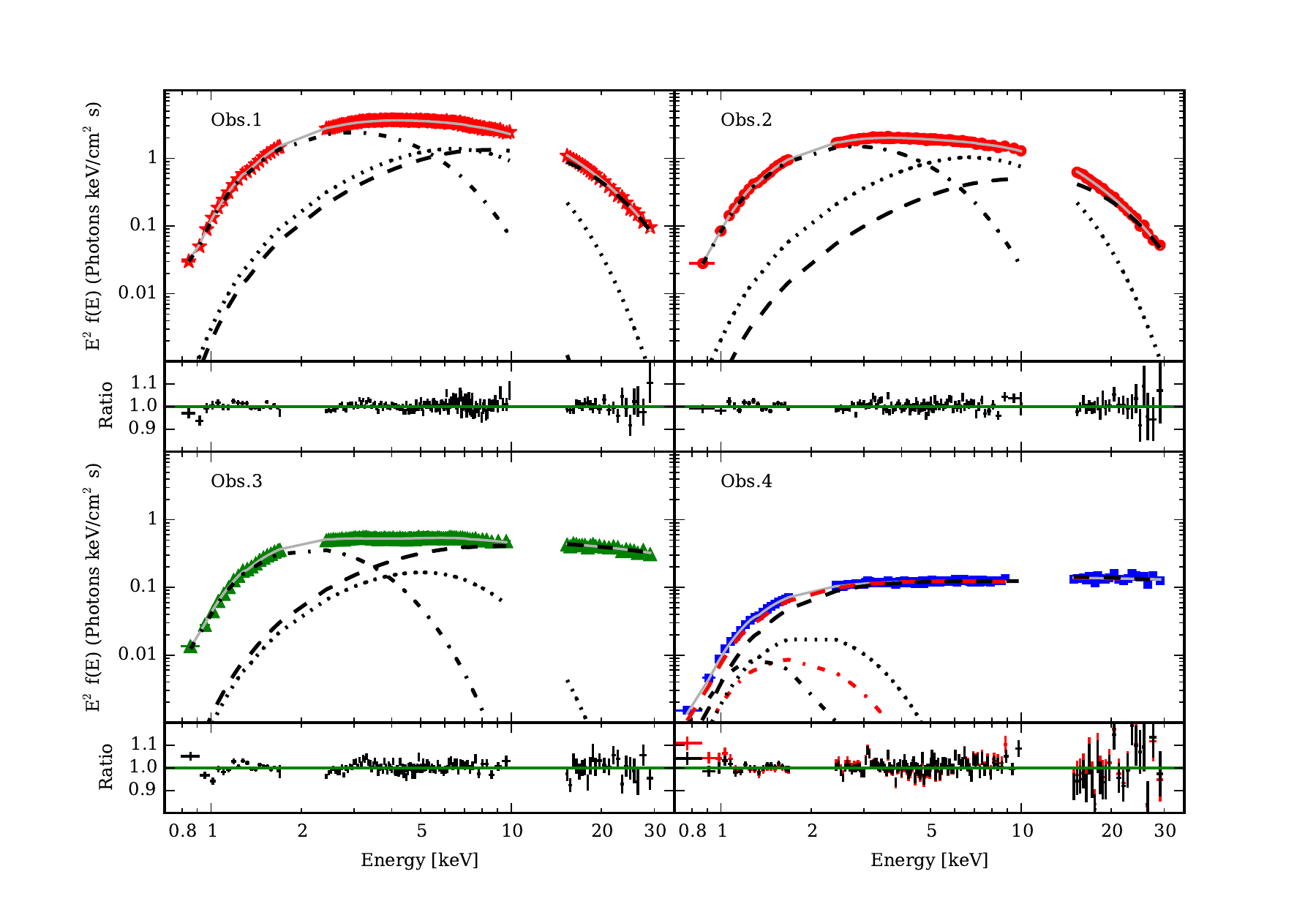}\\
\caption{Unfolded spectra and data-to-model ratio using the DISKBB+BBODYRAD+NTHCOMP model. \textbf{The full fit} is shown as a solid grey line, the disc component as a dot-dashed line, the blackbody component as a dotted line and the Comptonized component as a dashed line.  In the last panel the obs. 4 results obtained by using the simpler DISKBB+NTHCOMP model are overplotted in red. XIS data are re-binned in XSPEC for clarity.}
\label{fig:fit}
\end{center}
\end{figure*}


\subsection{Hard state observation}\label{subsec:HS}

We have successfully fitted the hard state observation (Obs. 4) with the 3-component model. We tried also the simpler 2-component approach, which includes the Comptonization component and only one of the thermal emissions (either MCD or BB). We find that the DISK+NTHCOMP model fits adequately the data with \chis = 1487 and 1403 dof . The corresponding spectral parameters are shown in the last column of Table \ref{tab:res} and the fit itself (red dashed line) in the last panel of Fig. \ref{fig:fit}. On the other hand, the BB+NTHCOMP model results in \chis of 1507 and soft residuals are observed. Statistically speaking, the former approach is consistent with the null hypothesis at 5 per cent significance level (p=0.057), while it is rejected by the latter one.  
Therefore, we find that both the 3-component model and a simplified version of it (DISK+NTHCOMP) are able to reproduce the data.  Even if degeneracy is obviously an issue, we note that an F-test indicates that the 3-component model (\chis of 1445 with 1400 dof.) is significantly better. In addition, the obtained spectral parameters follow the same trend than for  the previous 3 observations, suggesting that this modelling provides a realistic description of the spectral evolution of the system.

\begin{figure}
\begin{center}
\includegraphics[keepaspectratio,width=\columnwidth]{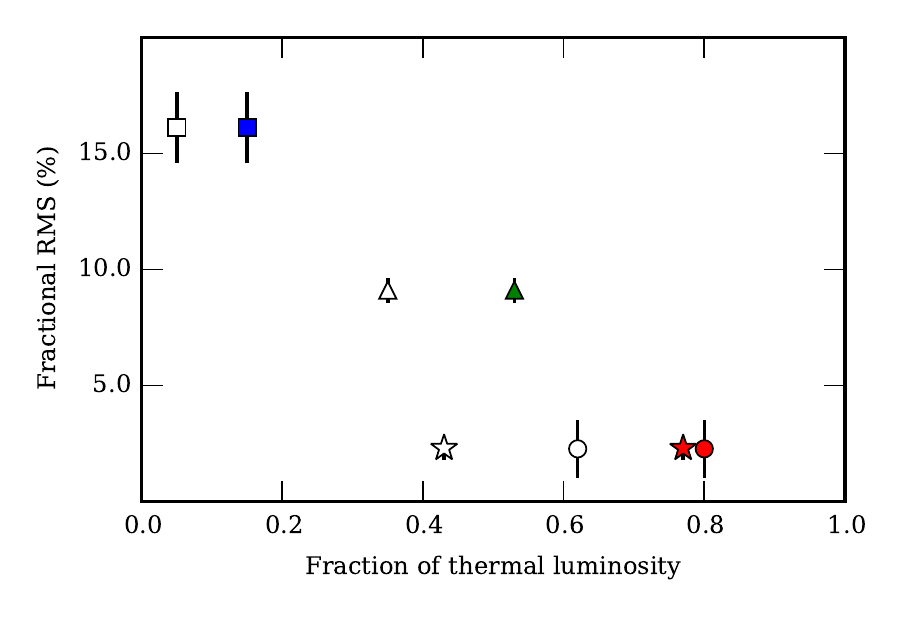}\\
\caption{RMS fraction versus fraction of thermal luminosity (DISKBB+BB). Filled and empty symbols correspond to 3-component model solutions assuming that the seed photons for the Comptonized emission arise from the NS surface and the accretion disc, respectively. }
\label{fig:Tfrac}
\end{center}
\end{figure}


\section{Discussion}\label{sec:Dissc} 

Model degeneration is a fundamental problem when studying the spectral evolution on NS-LMXBs. Here, we take advantage of the broad-spectral coverage and good spectral resolution of \suzaku\ to test the 3-component model proposed by \citetalias{Lin2007} as part of the \textit{hybrid model}. To this end, we have used 4 observations of the NS-LMXB \fourU\ taken during its 2010 outburst decay, covering the soft-to-hard transition (two in the soft state [SS], one in the intermediate state [IS] and one in the hard state [HS]).

Our 3-component model combines MCD, BB and NTHCOMP, which account for the emission from the accretion disc and NS surface, and the Comptonization contribution, respectively. Using this model we have successfully fit all four observations, obtaining spectral parameters which are within typical ranges from previous studies on \fourU\  \citep[e.g.][]{Gierlinski2002, Lin2007, Degenaar2015} and other NS LMXBs \citep[e.g.][]{Done2002, Lin2009, Takahashi2011, Sakurai2012}.

The evolution of each component through the different accretion states could be interpreted according to the following picture:
In the soft state, most of the emission is produced by the the NS surface and the accretion disc (which probably extends close to the NS surface) and only a little amount of radiation ($\sim 20$\%) is produced by Comptonization of seed photons emitted by these components. As accretion drops, the disc temperature and its associated luminosity decrease, which might be a consequence of disc truncation. Following this drop in accretion rate, there is a subsequent decrease on the NS surface temperature and luminosity. Thus, the Comptonization emission starts to be dominant, and eventually becomes the main source of emission in the hard state. This simplistic picture is similar to what is seeing in BH-LMXBs, with the difference that for NS systems an additional soft component is required to account for the NS surface. The presence of this extra thermal radiation should have an important role in producing the Comptonized emission. 

\subsection{The nature of the Comptonized emission}\label{comp}

So far we have discussed results obtained by assuming that the main source of seed photons for the Comptonization process is the NS surface (or boundary layer; i.e. we coupled $KT_{\rm seed}$ with $KT_{\rm bb}$). However, good fits are also obtained when assuming that the disc is the main source of seed photons (fifth column in Table \ref{tab:res}). Indeed, the obtained temperatures and photon indices are in the range expected for NS-LMXBs.   
Fig. \ref{fig:Tfrac} shows the evolution of the fractional rms as function of the thermal luminosity fraction for both possible seed photon solutions. When comparing the two approaches, the first striking thing is that the thermal fraction drops to $\sim$40 per cent in obs. 1, versus $\sim$80 per cent if the NS surface solution is adopted. The former value is much lower than that observed in BHs at the same variability levels (note that variability is generally observed to trace the Comptonization fraction;  e.g. \citealt{Remillard2006}). Moreover, the values obtained for the soft states observations (obs. 1 and 2) differ significantly  from each other when the disc approach is used. This is at odds with the similar hardness and rms values measured, both quantities being model independent and typically used to track the evolution of the accretion flow (Fig. \ref{fig:RXTE}). Nevertheless, we note that, as usual, the complexity of the nature exceeds any spectral fitting and probably both NS surface and disc photons are Comptonized. Thus, it is important to bear in mind that our model is only able to favour one components as the dominant source of seed photons. On the other hand, from Fig. \ref{fig:Tfrac} we cannot rule out the disc as the main source of seed photons for obs. 3 and 4. Indeed, our spectral modelling with 2-components strongly suggest that, if anything, the disc component might provide the most relevant thermal contribution in the hard state.

\subsection{Black holes comparative}\label{BHcomp}

One of the main advantages of the 3-component model is the straightforward comparison that it provides with BH-LMXBs, which show only one thermal component (i.e. the disc).  Our soft state and intermediate temperatures are consistent with values typically observed in BH systems \citep[e.g.,][]{Munoz-Darias2013} in agreement with findings by \citetalias{Lin2007}. However, the big advantage of our \suzaku -based study is the ability to study the thermal components in the hard state, when these are much weaker. \citet{Plant2015} modelled hard state observations of GX~339-4, a BH transient that is seen at an intermediate orbital inclination, as it is probably the case of \fourU. They find $KT_{\rm diskbb}$ in the range 0.15--0.28 in remarkable agreement with our work ($KT_{\rm diskbb}=$ 0.23	$\pm$0.02). In the same way, and using \suzaku\ observations, \citet{Shidatsu2014} find a similar range of disc temperatures in the hard state of H1743--322. 

Finally, we note that we do not see a strong evolution in our photon indices, which decrease from 2.3 to 2. While the former value is consistent with BH soft states ($\Gamma\sim$ 2--3), values $\Gamma\lesssim 2$ are more typical for the hard state. We note, however, that our hard state observation is one of the softest of \fourU\ in this state, as significantly larger rms and hardness values have been observed (see Fig. \ref{fig:RXTE}).  

\section{Conclusions}

We have performed a detailed study of the spectral evolution of the neutron star X-ray transient \fourU\ during the decay of its 2010 outburst. Our 0.8--30 keV \suzaku\ observations covered soft, intermediate and hard states epochs. We find that the 3-component model provides an excellent description of the outburst evolution. The inferred spectral parameters evolve in agreement with expectations, fulfilling several physical criteria, and reporting values consistent with those seen in other NS sources and in BH transients, whose spectral evolution is thought to be less complex. In the hard state, where the thermal emission is weaker and less energetic, the 3-component model is statistically preferred over simpler solutions. Likewise, we find that the disc component, typically observed in BHs when low energy coverage is available, is in any case required. Future works including more observations (e.g. at different stages of the hard state) and a variety of sources should be able to further extend this study. To this end,  we note that high-quality coverage at low energies is a desired feature, especially at low fluxes.

\section*{Acknowledgements}

MAP was funded by the International Research Fellowship program of the Japan Society for the Promotion of Science (PE15024). MAP and TMD acknowledge the hospitality of the Kyoto University, where part of this work was carried out. MAP acknowledge the hospitality on her visit to RIKEN, and thanks the valuable discussion with K. Makishima, T. Mihara and the other members of the MAXI team. MAP's research is funded under the Juan de la Cierva Fellowship Programme of the Ministry of Science and Innovation (MINECO) of Spain. MAP and TMD acknowledge support by the Spanish MINECO grant AYA2013-42627. MS acknowledges support by the Special Postdoctoral Researchers Program at RIKEN. This research has made use of data obtained from the Suzaku satellite, a collaborative mission between the space agencies of Japan (JAXA) and the USA (NASA).




\bibliographystyle{mnras}
\bibliography{/Users/montserrat/Documents/Mendeley/BibTex/library}

\bsp	
\label{lastpage}
\end{document}